\documentclass{PoS}

\let\oldbibliography\thebibliography
\renewcommand{\thebibliography}[1]{\oldbibliography{#1}
\setlength{\itemsep}{0pt}}

\title{New Venues in Formation and Detection for Primordial Black Hole Dark Matter}

\ShortTitle{New Venues for Primordial Black Holes}

\author{\speaker{Volodymyr Takhistov} \\
        Department of Physics and Astronomy, University of California, Los Angeles\\
 Los Angeles, CA 90095-1547, USA\\
        E-mail: \email{vtakhist@physics.ucla.edu}}

\abstract{Primordial black holes (PBHs) are not as exotic as once thought and constitute a compelling non-particle dark matter (DM) candidate. We present a novel general PBH formation mechanism from scalar field fragmentation, which does not suffer from the inflaton potential fine-tuning that plagues many of the standard PBH formation models. We discuss how interactions of compact stars with very small sub-lunar/asteroid-size PBHs, which reside in the open window of parameter space where PBHs can constitute all of DM, allow for a slew of new astrophysical signatures that could shed light on PBH DM and are particularly interesting in the era of multi-messenger astronomy.}

\FullConference{36th International Cosmic Ray Conference -ICRC2019-\\
		July 24th - August 1st, 2019\\
		Madison, WI, U.S.A.}

\begin{document}

\section{Introduction}

Standard astrophysical black holes arise at the end of stellar evolution as a result of star collapse, occurring at relatively late cosmological times after the star formation epoch. In contrast, black holes could also appear in the early Universe~\cite{Zeldovich:1967,Hawking:1971ei,Carr:1974nx}. These ``primordial'' black holes (PBHs) can constitute part or all of the dark matter (DM) (see Ref.~\cite{Carr:2016drx,Sasaki:2018dmp,Khlopov:2008qy} for review). Such non-particle DM candidates are particularly intriguing in light of lack of any convincing signals from particle DM searches, despite extensive experimental efforts~\cite{Liu:2017drf}. Further, PBHs have been associated with the recently observed~(e.g. \cite{Abbott:2016blz,Abbott:2016nmj,Abbott:2017vtc}) gravitational waves
(e.g.~\cite{Nakamura:1997sm,Clesse:2015wea,Bird:2016dcv,Sasaki:2016jop}) and can assist with resolution of major puzzles in astronomy, such as seeding formation of supermassive black holes~(e.g.~\cite{Kawasaki:2012kn}).

Primordial black holes are more generic than once thought and can form in a large variety of scenarios associated with physics beyond the Standard Model (BSM), including different models of inflation, axion-curvaton models (e.g.~\cite{Kawasaki:2012wr}), baryogenesis~\cite{Kawasaki:2019iis}, phase transitions as well as collapse of cosmic strings~(e.g.~\cite{Polnarev:1988dh}), among others (see Ref.~\cite{Carr:2016drx,Sasaki:2018dmp,Khlopov:2008qy} for an overview). In a typical setup, BHs form in the early Universe when an $\delta \sim \mathrm{O}(1)$ overdensity perturbation enters the horizon after the end of inflation and subsequently collapses. This allows to simply estimate the size of the resulting BH to be $M_{\rm PBH} \sim M_{\rm H} \sim 10^{15} (t/10^{-23}$~s)~g, where $M_{\rm H}$ is the horizon mass at the formation time $t$. Interestingly, PBHs can in principle already occur within ``standard cosmology'', albeit with an extremely suppressed probability. The usual perturbations arising from inflation are nearly scale-invariant and must agree with measurements from Cosmic Microwave Background (CMB) observations~\cite{Aghanim:2018eyx} on large scales. Hence, in order to obtain a sizable perturbation on small (PBH)-scales a significant degree of inflaton potential fine-tuning is expected within typical models of PBH formation~(e.g.~\cite{Kawasaki:2018daf,Germani:2018jgr}).~Formation associated with meta-stability of the Standard Model Higgs has been also proposed~\cite{Espinosa:2017sgp}, however, a full model with this mechanism is more complicated. Many open questions associated with PBH formation remain, such as clustering (e.g.~\cite{Ali-Haimoud:2018dau}).

We present a novel general mechanism of PBH formation based on scalar field fragmentation, following Ref.~\cite{Cotner:2016cvr,Cotner:2017tir,Cotner:2018vug,Cotner:2019ykd}. Aside from existence of the recently observed Higgs boson, scalar fields generically appear in BSM constructions. In particular, presence of moduli and axion fields are a defining feature  of models based on extra dimensions and string theories~\cite{Arvanitaki:2009fg}. Analogously, phenomenological models based on supersymmetry (SUSY) are accompanied by a slew of complex scalar fields with nearly vanishing potentials, i.e.~``flat directions''~\cite{Gherghetta:1995dv}. In the early Universe and in the presence of self-interactions, such scalar fields can often undergo fragmentation into solitonic lumps, Q-balls~\cite{Coleman:1985ki} or oscillons~(e.g.~\cite{Kasuya:2002zs}). Since the process is stochastic, some of the resulting sub-horizon overdensities arising from collection of field lumps could collapse and form BHs. As the relevant perturbations are not related to inflation, this mechanism does not suffer from the usual fine-tuning of the inflaton potential associated with many of the standard PBH-formation scenarios. Within this general setup, PBH formation can occur either before or after reheating and originate from inflaton or some other, unrelated, spectator field. Interestingly, unlike the usual radiation-era formation models, PBHs from field fragmentation are formed during matter-dominated era and could possess large spins.

Formed primordial black holes can contribute to DM if their size is $M_{\rm PBH} \gtrsim 10^{15}$~g, sufficient to survive Hawking evaporation until present day\footnote{Evaporating PBHs significantly contributing to DM could also lead to additional constraints~(e.g.~\cite{DeRocco:2019fjq,Laha:2019ssq}).}. While it is difficult to realize all of DM to reside in PBHs over most of the PBH parameter space, there remains an unconstrained open window in the $\sim 10^{17} - 10^{22}$~g sub-lunar/asteroid-size mass range. Such very small PBHs can be captured by compact stars, neutron stars (NS) and white dwarfs (WD), in DM-rich environments~\cite{Capela:2013yf}. As captured PBHs evolve consuming the host star, they will eventually destroy it. We discuss, following Ref.~\cite{Fuller:2017uyd,Takhistov:2017bpt,Takhistov:2017nmt}, how these PBH-star systems can result in a variety of novel astrophysical signatures, including solar-mass black holes as well as ``orphaned'' kilonovae or gamma-ray bursts (GRBs) without associated merger gravity waves. This allows for potential novel insights into the open PBH parameter space window that is difficult to probe otherwise. Further, PBH-star systems might also help with resolving some of the long-standing astronomical puzzles, such as the origin of $r$-process nucleosynthesis~\cite{Fuller:2017uyd}.

\section{Primordial black holes from scalar field lumps}

\subsection{Scalar field evolution and fragmentation in the early Universe}

Consider a light complex scalar field $\phi$ in a potential $V(\phi)$, charged under some global $\mathrm{U}(1)$ symmetry, which does not dominate the energy-density in the early Universe. This setup is typical of supersymmetric models, where many of the flat directions are found to be charged under $\mathrm{U}(1)$ baryon or lepton number~\cite{Gherghetta:1995dv}. Analogously to the Standard Model Higgs, such a spectator field will experience random quantum jumps during the inflationary (de Sitter) phase~\cite{Enqvist:2013kaa}. The field, off-set from its minimum of $\phi = 0$ to $\phi_0$, can hence develop a large vacuum expectation value (VEV) as determined by $V(\phi_0) \sim H_I^4$. Here, $H_I$ is the Hubble parameter value associated with the scale of inflation $\Lambda_I$ via $H_I = \sqrt{8 \pi/3} (\Lambda_I^2/M_{\rm pl})$,
where $M_{\rm pl} \simeq 2.45 \times 10^{18}$ GeV is the reduced Planck mass. 

After inflation, the scalar field will roll back down relaxing to the potential minimum and coherently oscillate. In the presence of self-interactions, growing instabilities can develop in the field's oscillation modes. Parametrizing the complex field as a real field with a rotating phase $\phi = R(x,t) e^{i \Omega(x,t)}$, it can be shown through analysis of dispersion relations that unstable growing oscillation modes appear under simple and very general condition ~\cite{Kusenko:1997si}
\begin{equation}
    V''(R) - \dot{\Omega} < 0~.
\end{equation}
For the special case of $\dot{\Omega} = 0$, the above reduces to the popular ``tachyonic resonance''. Growing instability modes will eventually result in scalar condensate fragmenting into solitonic lumps, Q-balls~\cite{Coleman:1985ki}. The evolution of instabilities associated with the scalar field self-interactions are analogous to gravitational Jeans instability of self-gravitating gas, often appearing in astrophysics and cosmology. The attractive self-interactions of
gravity make homogeneous distribution of particles unstable with respect to formation
of dense matter clumps, leading to collapse of gas clouds and subsequent star formation. Another way to see that in the presence of self-interactions lump formation will be generally expected is by looking at the pressure($p$)-density($\rho$) relations. In particular, for field potential effectively rising slower than quadratic, i.e.~$V(\phi) \sim |\phi|^n$ with $n < 2$, the pressure of a system $p \propto (- \rho)$ is negative and hence leads to condensate collapse~\cite{Turner:1983he}.

Field perturbations $\delta \phi$ associated with the instability and growing faster than the expansion rate of the Universe will eventually enter the non-linear regime (i.e. $\delta \phi \gtrsim \overline{\phi}$, where $\overline{\phi}$ is the average background field value) and can subsequently fragment. Existence of spherical solitonic Q-ball configurations in the spectrum of a theory with self-interacting scalar fields charged under $\mathrm{U}(1)$ can be rigorously shown from energetic considerations~\cite{Coleman:1985ki}. Long-term stability of Q-balls is ensured by charge conservation. 
The size of the resulting Q-ball lumps can be determined by analyzing growing oscillation modes that become non-linear the earliest, denoted $k_{\rm nl}$. From numerical simulations~\cite{Kusenko:1997si,Kasuya:2000wx,Multamaki:2002hv} the size of a typical field fragment $R_Q$ constitutes a few percent of the horizon size at fragmentation
\begin{equation}
    R_Q \sim k_{\rm nl}^{-1} \sim f_Q H^{-1}~,
\end{equation}
where $f_Q \sim 10^{-1} - 10^{-2}$. The resulting number density of solitons is then approximately $\overline{n} \sim (k_{\rm nl}/2 \pi)^3 \sim 10 - 10^6$. Generally, the Q-ball mass and radius are given by
\begin{equation}
    M_Q = \Lambda Q^{\alpha}~~~~~,~~~~~R_Q = \frac{Q^{\beta}}{\Lambda}
\end{equation}
where $\Lambda$ is a scale associated with scalar field potential, with $0 < \alpha$ and $\beta < 1$ being model-dependent parameters. For gauge-mediated SUSY, $\Lambda = M_{\rm SUSY}$ parametrizes the SUSY breaking scale and $\alpha = 3/4$, $\beta = 1/4$~\cite{Kusenko:1997si}. While Q-balls are stable with respect to decays into $\phi$-quanta, they could decay e.g. due to higher-dimensional operators that break the $\mathrm{U}(1)$ symmetry. These effects can be phenomenologically parametrized by the Q-ball lifetime $\tau_Q = 1/\Gamma_Q$, with $\Gamma_Q$ being the total decay width.

The discussion above can be readily extended to the case of real scalar field fragmenting into pseudo-solitonic lumps, oscillons~(e.g.~\cite{Kasuya:2002zs}). While the stability of oscillons is not guaranteed by a $\mathrm{U}(1)$ symmetry, they could be long lived due to an approximate adiabatic invariant that can be related to particle number~\cite{Kasuya:2002zs}. In fact, oscillons can be identified with the real projection of the Q-ball solution in non-relativistic field theory limit~\cite{Mukaida:2014oza}.

\subsection{Primordial black hole formation}

After the end of inflation, inflaton decays onset radiation-dominated era. Subsequent fragmentation of a spectator scalar field results in population of solitonic lumps of matter. Since matter redshifts as $a^{-3}$, compared to radiation $a^{-4}$, with $a$ being the cosmological scale factor, at some later time Q-balls come to matter-dominate the energy-density of the Universe. As the fragmentation process is stochastic and the resulting lumps are sizable, some sub-horizon regions could contain sufficiently large overdensities that will subsequently collapse and form black holes.

At fragmentation time $t_f$, the resulting PBH-mass function can be obtained via~\cite{Cotner:2016cvr,Cotner:2017tir,Cotner:2018vug,Cotner:2019ykd}
\begin{equation} \label{eq:pbhcol}
\frac{d\langle{\rho_{\rm PBH}\rangle}}{dM}\Big|_{t_f}  = \int \frac{dV}{V^2} P(M|V)\, B(M,V) M~,
\end{equation}
where $P(M|V)$ is the probability of finding a soliton cluster of mass $M$ within volume $V$ and $B(M,V)$ is the probability of cluster of mass $M$ to collapse to a BH. Assuming the most general model-independent and uncorrelated distribution of soliton lumps, we take $P(M|V)$ to be Poisson.
The BH collapse condition is approximately of the form $B(M,V) \sim K \theta[\delta_0 - \delta_c]$, with the step function $\theta$ selecting original soliton cluster overdensities $\delta_0$  that lie above a critical threshold for collapse $\delta_c \sim \mathrm{O}(1)$. The phenomenological constant prefactor $K$ parametrizes any additional complex effects that play a role in the collapsing set of solitons, including angular momentum, anisotropy, etc. (see Ref.~\cite{Polnarev:1986bi,Harada:2016mhb,Harada:2017fjm} for related treatment of collapsing dust in matter-dominated era). Since Q-ball charge is conserved, there is an additional corresponding condition is imposed on resulting fragments.

In order to obtain the present-day contribution, results of Eq.~(\ref{eq:pbhcol}) must be redshifted through the initial radiation-dominated and subsequent Q-ball matter-dominated eras. The Q-ball matter-dominated era starts at $t_Q$ and finishes at $t_R$, when the Q-balls that have not collapsed to BHs decay away and reheat the Universe to temperature $T_R$. In order not to spoil the Big Bang Nucleosynthesis, we impose $T_R \gtrsim 5$ MeV. The evolution through above cosmological phases is captured in a multiplicative scale factor
	\begin{equation} \label{eq:scalefacqball}
	a_t = \left(\frac{t_Q}{t_f}\right)^{1/2} \left(\frac{t_R}{t_Q}\right)^{2/3} \left(\frac{g_{\ast}(T_R)}{g_{\ast}(T_0)}\right)^{1/4} \frac{T_R}{T_0}~,
	\end{equation}
where $T_0 = 2.7$ K is the present-day temperature and $g_{\ast}(T)$ denotes the relevant number
of temperature-dependent relativistic degrees of freedom.

The final resulting differential fraction of
dark matter in PBHs can be obtained from
\begin{equation}
    \frac{df_{\rm DM}}{dM} = \frac{1}{\rho_{\rm DM}} \frac{1}{a_t^3} \frac{d\langle{\rho_{\rm PBH}\rangle}}{dM}\Big|_{t_f}~,
\end{equation}
where $\rho_{\rm DM}$ the observed DM abundance. The above analysis is general and readily applicable to oscillons\footnote{Some related oscillon simulations have been recently initiated in Ref.~\cite{Lozanov:2019ylm}. We stress, however, that it could be a difficult task for simulations to properly
capture our proposed PBH formation scenario from the very rare occurrences with all the relevant
effects taken into account.},
which could originate directly from the inflaton fragmentation~\cite{Cotner:2018vug}, or to any other possible solitonic configurations with any number of conserved quantities beyond just the scalar field charge. 

\begin{figure*}[htb]
\begin{minipage}[b]{0.5\textwidth}
\centering
\includegraphics[width=\textwidth]{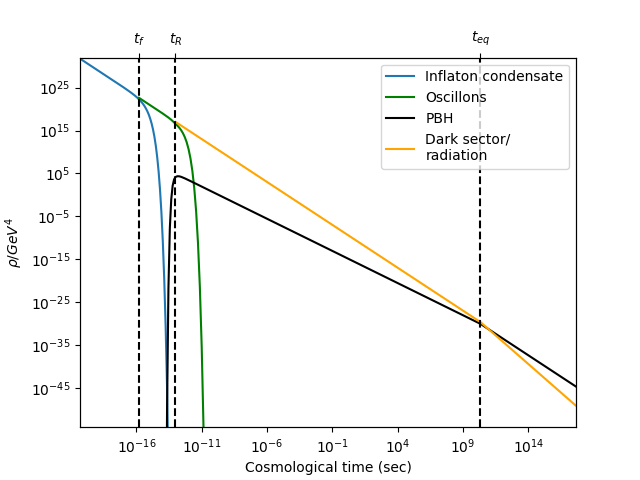}
\end{minipage}
\hspace{-1.5em}
\begin{minipage}[b]{0.5\textwidth}
\centering
\includegraphics[width=1.125\linewidth]{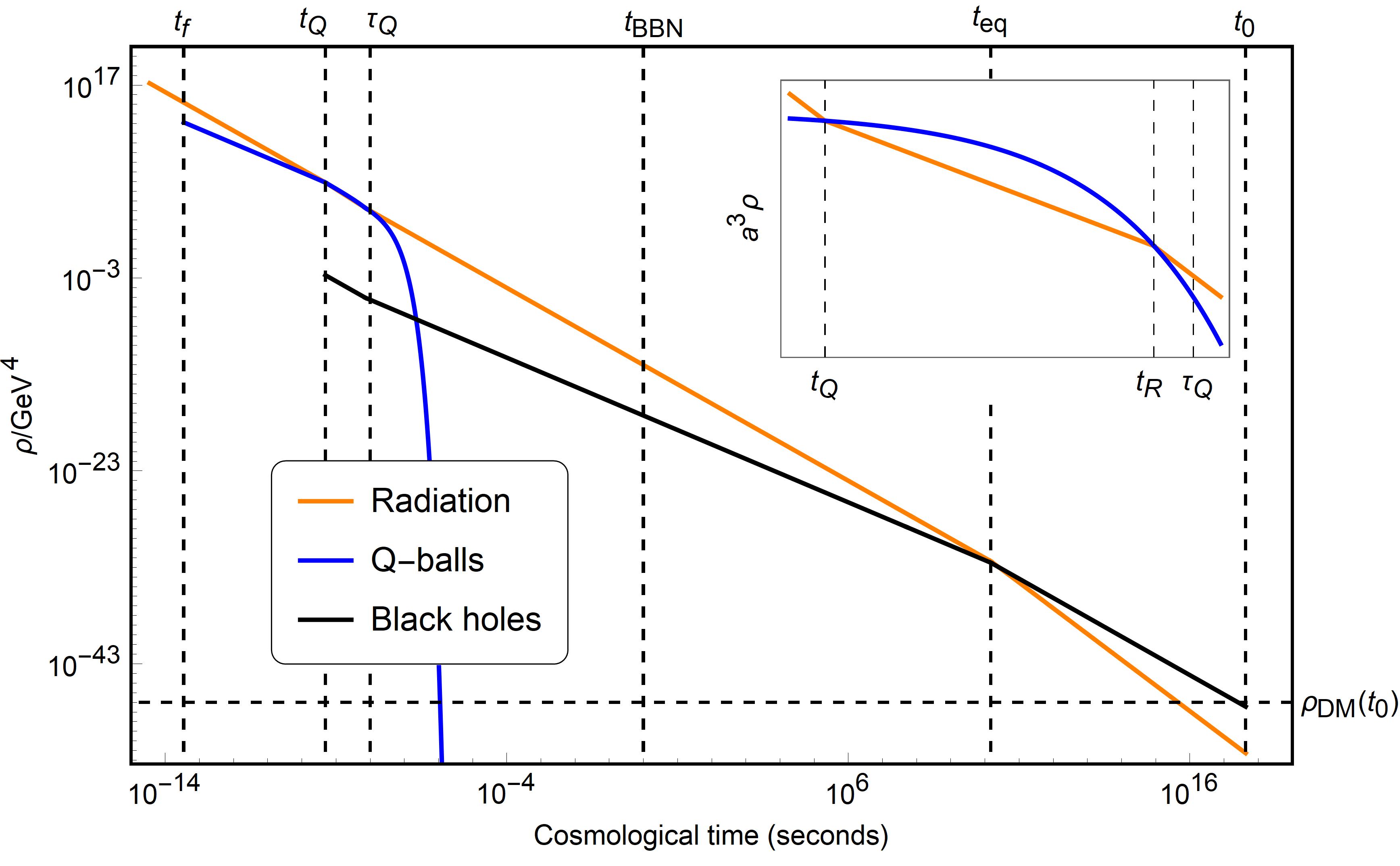}
\end{minipage}
		\caption{Two examples of a typical timeline. [left panel] In the model from Ref.~\cite{Cotner:2018vug},  oscillon scalar lumps formed directly from the inflaton dominate the energy density for a limited time creating a temporary matter-dominated era, during which  PBHs form. [right panel] In the case of a spectator field~\cite{Cotner:2016cvr,Cotner:2017tir}, it can fragment into Q-balls during the radiation-dominated era.  Since the gas of Q-balls has the energy density that scales as matter, it comes to dominate at time $t_Q$, creating a matter dominated era that lasts until the Q-balls decay at time $\tau_Q$.~PBHs form during this matter-dominated era. 
	 }
		\label{fig:timeline}
\end{figure*}

In Fig.~\ref{fig:timeline} we illustrate the typical cosmological timelines for PBHs forming from the inflaton fragmenting into oscillons (left panel), without an intermediate radiation-dominated era (i.e. $t_f = t_Q$), and from a spectator scalar field fragmenting into Q-balls (right panel).

\section{Compact stars as primordial black hole laboratories}

 There remains an open parameter-space window for sub-lunar/asteroid-size PBHs of $\sim 10^{17} - 10^{22}$ g mass to constitute all of DM. In particular, recent re-analyses of the previously suggested constraints for this parameter range originating from neutron star capture in globular clusters~\cite{Capela:2013yf}, femtolensing~\cite{Barnacka:2012bm} and white dwarf abundance~\cite{Graham:2015apa} found them to not be as effective\footnote{Constraints from GRB femtolensing are affected by extended nature of sources as well as wave optics effects~\cite{Katz:2018zrn}, reducing their applicability. Similarly, constraints from HSC~\cite{Niikura:2017zjd} do
not apply below $\lesssim 10^{23}$ g, since associated lensing magnification is strongly suppressed when the size of BH becomes smaller than the relevant wavelength of light. It is also difficult to constrain small PBHs constituting DM by analyzing NS stability within globular clusters, as observations of  these systems show no evidence of significant DM abundance~\cite{Bradford:2011aq,Sollima:2011nb}. Further, a more detailed study including simulations of WD heating and destruction by transiting small PBHs found the associated constraints to also not be very robust~\cite{Montero-Camacho:2019jte}.}.

PBHs whose masses lie within the open parameter window of $\sim 10^{17} - 10^{22}$ g and that constitute  a significant fraction of DM can efficiently interact with compact stars within DM rich environments, such as Galactic Center (GC)  or Ultra-faint Dwarf Spheroidal (UFD) galaxies. A small PBH can become gravitationally captured by a neutron star or a white dwarf if it loses sufficient energy through dynamical friction and accretion as it passes through the star~\cite{Capela:2013yf}. For typical parameter values and optimistic DM density consistent with observations, the base capture rate of PBHs on NS can be estimated to be $F_0^{\rm GC} \simeq 10^{-11}/$yr within the GC and an order better for UFDs~\cite{Fuller:2017uyd}. The full capture rate depends on the PBH fraction constituting DM as $F = (\Omega_{\rm PBH}/\Omega_{\rm DM}) F_0$, with $\Omega$ denoting the respective abundance. The total number of captured PBHs within time $t$ is $F t$. Due to significantly lower density, the capture rate on WDs is smaller by several orders compared to NS. After capture, PBH will settle inside the star and consume it through accretion~\cite{Kouvaris:2013kra}.~For a typical NS, with mass of $M_{\rm NS} \sim 1.5 M_{\odot}$ and radius of $R_{\rm NS} \sim 10$ km, the time for a captured PBH to settle within star is $t_{\rm set}^{\rm NS} \simeq  9.5 \times 10^3 (M_{\rm PBH}/10^{-11} M_{\odot})^{-3/2}$ yrs.~After settling, the time for BH to consume the star form the inside is 
$t_{\rm con}^{\rm NS} \simeq 5.3 \times 10^{-3} (10^{-11} M_{\odot}/M_{\rm PBH})$ yrs. The resulting star's lifetime is determined by the combination of above time-scales.

Anticipating the role of angular momentum, it is particularly interesting to look at interactions of PBHs with pulsars that are rotating extremely rapidly with a millisecond period~\cite{Fuller:2017uyd}. It can be shown that within the age of the Galaxy of $\sim 10^{10}$ yrs, up to $\sim 10\%$ of $\sim 10^7$ millisecond pulsars (MSPs) residing within GC would have been consumed by PBHs if they are to constitute a significant fraction of DM. This is consistent with the ``missing pulsar problem'', denoting under-abundance of observed pulsars in the center of the Galaxy compared to predictions from population synthesis models~\cite{Dexter:2013xga}. A recently discovered very young magnetar located right near the GC is consistent with our the scenario outlined above and its unusual emission activity~\cite{Zelati:2015vya} might be a sign of a PBH destruction in progress. Assuming that angular momentum, originally associated with material infalling into the growing by accretion central black hole within the star, is efficiently transferred to the star's outer layers, analytic estimates predict that more than $\sim 0.1 M_{\odot}$ of neutron-rich material can be ejected~\cite{Fuller:2017uyd}. Heuristically, MSPs are already rotating near mass-shedding limit and conservation of angular momentum during contraction of the star as it is being consumed by PBH will result in spin up. Hence, with further additional angular momentum transferred from within, matter on the star's surface will start to exceed the escape velocity.

The material ejected as a result of PBH-NS destruction is neutron rich and can be an excellent source of $r$-process nucleosynthesis, which is a key astrophysical process for production of heavy elements such as gold and uranium~\cite{Fuller:2017uyd}.
Here, neutrons are rapidly captured on a seed nuclei before they can decay, allowing to build up an element with a large atomic number. While supernovae and neutron-star mergers have been historically suggested as the beacons of $r$-process element production~(e.g.~\cite{Eichler:1989ve}), a definitive confirmation of production sites is still lacking.~Around $\sim 10^4~M_{\odot}$ of $r$-process material is observed within Milky Way. Recently, it has been noted that 1 out of 10 UFDs shows strong $r$-process abundance, consistent with a singular rare historic event~\cite{Ji:2015wzg}.~Heavy element production associated with PBHs destroying neutron stars can consistently explain abundance of $r$-process elements in Milky Way as well as UFDs simultaneously~\cite{Fuller:2017uyd}.
\begin{figure*}[htb]
\vspace{-1em}
\centering
\includegraphics[width=0.5\textwidth]{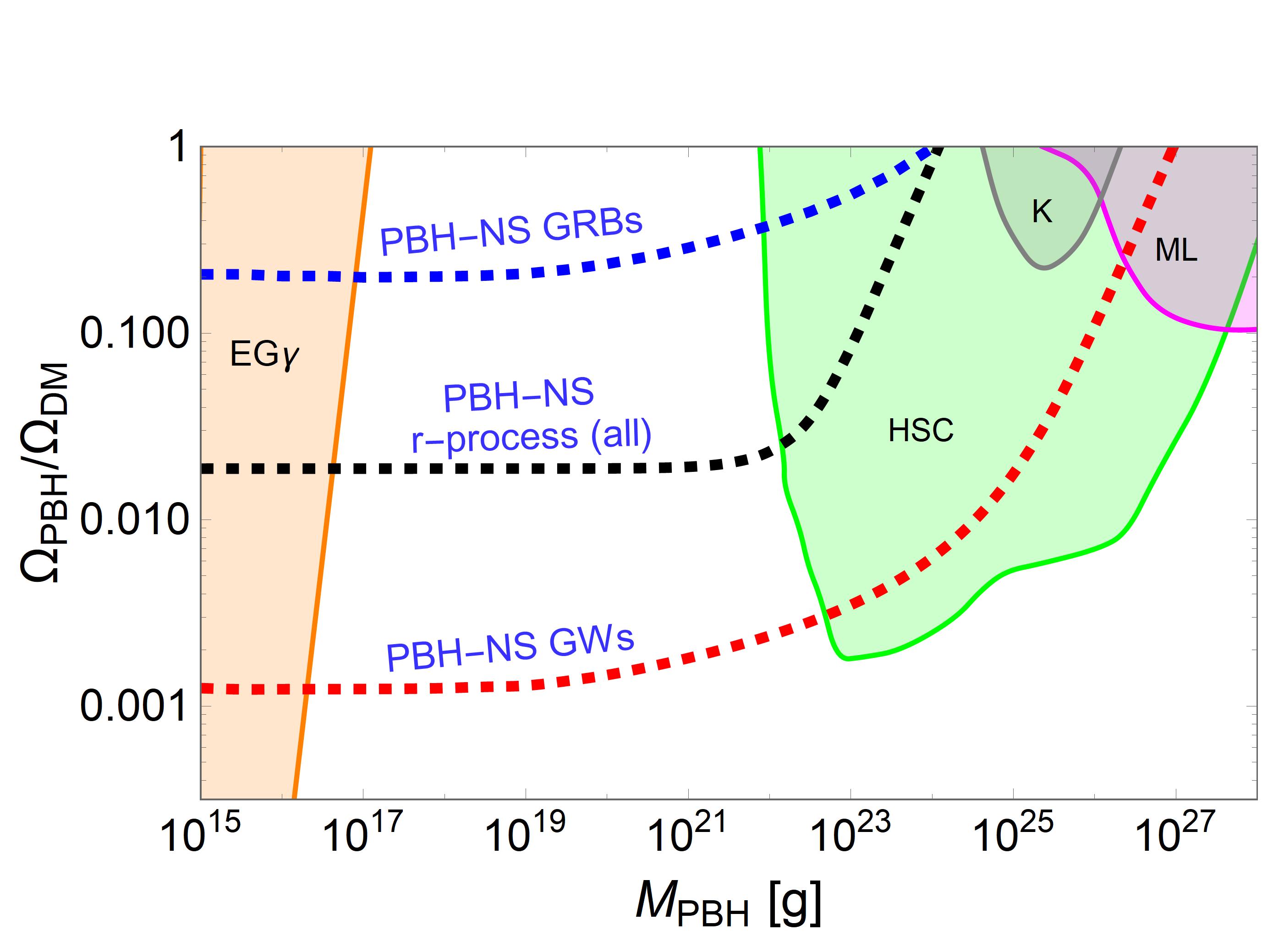}
		\caption{Signals from PBH-compact star interactions along with experimental constrains. [blue dashed] Allowed parameter space for PBH-NS gamma-ray bursts to significantly contribute to the observed positron excess at high energies~\cite{Takhistov:2017nmt}. [black dashed] Allowed parameter space where heavy element nucleosynthesis associated with PBH-NS systems can account for all of $r$-process element abundance in Milky Way as well as UFDs simultaneously~\cite{Fuller:2017uyd}. [red dashed] Allowed parameter space where gravity waves from transmuted binaries from PBH-NS interactions can be within observation reach of Advanced LIGO, assuming a lower signal-to-noise threshold than design~\cite{Takhistov:2017bpt}. Dashed lines correspond to the maximum reach for each signal, assuming the most optimistic input parameter choice for the fits. Observational constraints from extragalactic $\gamma$-rays from BH evaporation (EG$\gamma$)~\cite{Carr:2009jm},  Kepler star milli/microlensing (K)~\cite{Griest:2013aaa}, MACHO/EROS/ OGLE micro-lensing (ML)~\cite{Tisserand:2006zx} as well as Subaru Hyper Suprime-Cam micro-lensing (HSC)~\cite{Niikura:2017zjd} are shown.
	 }
		\label{fig:pbhnsfull}
\end{figure*}

In addition to the $r$-process nucleosynthesis, PBHs interacting with compact stars can result in a variety of novel astrophysical signatures~\cite{Fuller:2017uyd, Takhistov:2017bpt,Takhistov:2017nmt}\footnote{Similar type of signatures could also appear within the context of particle DM if sufficient amount of it accumulates within a compact star and leads to a black hole~(e.g.~\cite{Bramante:2017ulk}).}. We give a brief overview of some of these signals, displaying several of them in Fig.~\ref{fig:pbhnsfull} along with experimental constraints on PBH abundance in DM:
\begin{itemize}
    \item \textit{\underline{Orphan kilonovae}}:~~~ electromagnetic afterglow originating from expanding nuclear material ejecta will result in a ``kilonova''. However, in contrast to the standard kilonova signals as expected from neutron star mergers, kilonovae from PBH-NS systems will be without an associated gravity wave merger signal. 
    \item \textit{\underline{Non-repeating fast radio bursts (FRBs)}}:~~~a slew of FRB signals have been recently detected, with an unknown origin~\cite{Katz:2018xiu}. As neutron stars are highly magnetized, the energy associated with NS magnetic fields will be released during the final stages of PBH consumption of the star and can result in a non-repeating fast radio burst if even a small percentage is converted to radio signal~\cite{Fuller:2014rza}.
     \item \textit{\underline{Positrons and 511 keV radiation}}:~~~as the expanding ejecta is heated through nuclear interactions, thermal positrons will be produced and leaking out. These positrons will annihilate, resulting in a 511 keV emission. Crude estimates show that the strength of the expected 511 keV signal from PBH-NS systems is consistent with the observed long-standing 511 keV excess from the GC~(see Ref.~\cite{Prantzos:2010wi} for review). Further, recent observations~\cite{Siegert:2016ijv} of strong 511 keV emission signal originating from the same UFD (Reticulum II) showing enhanced $r$-process element abundance suggests a natural link between the two and both occurring due to a singular rare event. This can be readily incorporated within the PBH-NS scenario (or, alternatively, a neutron star merger~\cite{Fuller:2018ttb}).
    \item \textit{\underline{Solar-mass black holes}}:~~~as captured PBH consumes the host compact star (WD or NS), most of the original star material will end up swallowed by the BH. Hence, there will remain a population of $\sim 0.5-1.5 M_{\odot}$ solar-mass BH remnants, not expected from standard astrophysics. Further, a merging binary that originally contained a compact star that was later ``transmuted'' into a solar-mass BH could provide novel associated signals (e.g. double kilonova signals - one from merger and one from PBH-star interaction). Such a solar-mass BH  could be potentially distinguishable from a NS with future observations via e.g. higher order gravity wave effects.
    \item \textit{\underline{Orphan GRBs and solar-mass microquasars}}:~~~if some of the material from PBH-NS system forms an accretion disk surrounding the resulting solar-mass BH, this will be a natural setup for a short GRB. However, unlike a typical short GRB from neutron star mergers, such PBH-GRB will not have an accompanying merger gravity wave signal. 
    In the case of WDs, while accreting WDs can emit jets, due to large radius of the WD the jets are non-relativistic, with luminosity approximately scaling as $L \sim 1/R_{\rm star}$. A PBH captured by a WD will eventually consume it, resulting in a compact BH remnant. Due to significant decrease in radius of the accretor, the associated jet can suddenly become relativistic, resulting in a novel solar-mass microquasar. Interestingly, jet emission from a local population of PBH GRBs and solar-mass microquasars can contribute to the positron excess observed by several experiments, including PAMELA~\cite{Adriani:2013uda}, Fermi-LAT~\cite{FermiLAT:2011ab} and AMS-02~\cite{Aguilar:2013qda}.
\end{itemize} 
 
\section{Conclusions}
\label{sec:conclusions}

PBHs constitute a compelling non-particle DM candidate. We have presented a novel general PBH formation mechanism from scalar field fragmentation that can be implemented within a broad class of models and that avoids the issue of inflaton potential fine-tuning typically associated with many of the standard PBH formation scenarios. Interactions of compact stars with tiny PBHs, which reside in the open window of parameter space where PBHs can constitute all of DM,
can be a rich source of novel astrophysical signals that could shed light on PBH DM.

\subsubsection*{Acknowledgments}
We thank the organizers of ICRC-2019 for the opportunity to present our results. The work of
V.T. was supported by the U.S. Department of Energy (DOE) Grant No. DE-SC0009937.

\bibliography{pbhlib}
\addcontentsline{toc}{section}{Bibliography}
\bibliographystyle{JHEP}

\end{document}